\newcommand{\cl}{\centerline}
\documentstyle[12pt]{article}
\begin{document}
\def\d{{\rm d}}
\hfill{CCUTH-94-04}\par
\hfill{IP-ASTP-12-94}\par
\setlength{\textwidth}{5.0in}
\setlength{\textheight}{7.5in}
\setlength{\parskip}{0.0in}
\setlength{\baselineskip}{18.2pt}
\vfill
\cl{\large{{\bf Perturbative QCD Analysis of}}}\par
\cl{\large{{\bf $B$ Meson Decays}}}\par
\vskip 1.5cm
\cl{Hsiang-nan Li$^1$\ and Hoi-Lai Yu$^2$, }
\vskip 0.5cm
\cl{$^1$Department of Physics, National Chung-Cheng University,}
\cl{Chia-Yi, Taiwan, R.O.C.}
\vskip 0.3cm
\cl{$^2$Institute of Physics, Academia Sinica, Taipei, Taiwan, R.O.C.}\par
\vskip 1.0cm
\cl{\today }
\vskip 3.0 cm
\cl{\bf Abstract}

Resummation of large QCD radiative corrections, including leading and
next-to-leading logarithms, in pion electromagnetic form factor
is reviewed. Similar formalism is applied to exclusive processes involving
heavy mesons,
and leads to Sudakov suppression for the semi-leptonic decay
$B\to\pi l\nu$. It is found that, with the inclusion of Sudakov effects,
perturbative QCD analysis of this decay is possible for the energy
fraction of the pion above 0.3. By combining predictions from the
soft pion theorems, we estimate that the upper limit of
the KM matrix element $|V_{ub}|$ is roughly 0.003.

\vfill
\newpage

\cl{\large \bf 1. Introduction}
\vskip 0.5cm

It has been shown that
perturbative QCD (PQCD) is applicable to exclusive processes
such as elastic hadron form factors for energy scale higher than few
GeV \cite{LS}.
The enlargement of the range of the applicability from much higher
energies \cite{R} down to this low
scale is due to the inclusion of transverse momentum dependence
into factorization theorems. This dependence appears in an
exponential factor, which arises
from the resummation of large radiative corrections, and
leads to Sudakov suppression for
elastic scattering of isolated colored quarks. Detailed derivation
of Sudakov factors for hadron-hadron Landshoff scattering
refers to \cite{BS}. Similar expressions
have been obtained and employed in the
PQCD analysis of pion and proton form factors~\cite{LS,L,GB,BSW}, pion
Compton scattering \cite{CL} and other exclusive processes \cite{HSS}.
Resulting predictions from this modified version of factorization
theorems have been
examined and found to be dominated by perturbative contributions.

All the above analyses of large corrections
involves only light hadrons.
In this paper we shall extend the formalism to exclusive processes
containing both light and heavy mesons, such as $B$ meson decays,
and organize their Sudakov corrections to all orders.
An important work in the study of the standard model is to determine
the mixing angles in the Cabibbo-Kobayashi-Maskawa (KM) matrix.
The decay $K\to\pi l \nu$ contains the information of
the matrix element $|V_{us}|$, and chiral
symmetry provides a precise method to study this process \cite{LR}.
Similarly, $|V_{cb}|$ is determined
by exploring the $B\to D l \nu$ decay, for which
heavy quark symmetry is an appropriate tool \cite{N1}.
As to $|V_{ub}|$, which can be
measured reliably from the $B\to\pi l \nu$ decays \cite{IW},
neither of the above theories is proper.

Recently, an analysis of the semi-leptonic decay $B\to \pi l \nu$
based on the heavy quark effective theory
(HQET) has been performed by Burdman {\it et al.}
\cite{BLN}. They have determined the
normalization of form factors involved in the process in terms of the soft
pion relations, and given the ratio of these form factors to the
corresponding
ones in the $D\to \pi l \nu$ decay. However, explicit evaluation of this
process is not yet successful.
In this paper we shall show that, by incorporating Sudakov
effects, PQCD is indeed applicable to the semi-leptonic $B$ meson decays
as the pion carries away more than a quarter of its maximum energy.
The differential decay rate is then obtained.
By combining our predictions with those from
the soft pion theorems, which have been derived in the framework of
HQET \cite{BLN}, the total decay rate and
the branching ratio of this process are estimated.
Comparing this estimation with experimental data, we find that the upper
limit of the KM matrix element $|V_{ub}|$ is roughly 0.003.
PQCD is then able to complement HQET in the analysis of heavy meson decays.
Our results can be easily generalized to perturbative
analysis of other heavy-to-light transitions.

We find that factorization theorems are successful for this
semi-leptonic decay especially when the pion is energetic.
In this case non-perturbative approaches such as the soft pion theorems
and QCD sum rules are not applicable. The first
attempt to apply the modified perturbative formalism
to the analysis of the $B\to\pi l \nu$ decay has been made by Akhoury
{\it et al.} \cite{YS}. However, they did not consider the transverse
momentum dependence, and have obtained predictions which are drastically
different from ours.
As claimed in \cite{YS}, the hard gluon involved in the decay process,
with Sudakov effects taken into account,
is off-shell at most by an amount of
$1.4\Lambda_{\rm QCD}m_b$, $m_b$ being the $b$ quark mass. In our approach
the hard gluon is off-shell roughly by $8\Lambda_{\rm QCD}m_b$,
and therefore, the perturbative analysis is more reliable.

The resummation of Sudakov logarithms for the pion form factor
is reviewed in section 2.
The same formalism is then applied to semi-leptonic $B$ meson decays
in section 3, where
the full expression for the Sudakov factor including leading and
next-to-leading logarithms is given.
Section 4 contains numerical analysis of the
modified factorization formulas for the relevant form factors and the
differential decay rate of $B\to\pi l\nu$.
Section 5 is our conclusion.
\vskip 1.0cm

\cl{\large \bf 2. The Pion Form Factor}
\vskip 0.5cm

First, we review the modified factorization formula for a simple
light-to-light process, the pion electromagnetic form factor,
to lowest order of coupling constant $\alpha_s$, which is expressed as
the convolution of wave functions and a hard scattering amplitude.
We then investigate radiative corrections to this formula,
and explain how they are absorbed into the above convolution factors.
The first step is to find out the leading momentum regions of
radiative corrections,
from which important contributions to loop integrals arise.
There are two
types of important contributions: collinear, when the loop momentum
is parallel to the incoming or outgoing pion momentum, and soft, when
the loop momentum is much smaller than the momentum transfer $Q^2$ of the
process. Here, $Q^2$ is assumed to be large and
serves as an ultraviolate cutoff of loop integrals.
We associate small amount of transverse momenta
${\bf k}_T$ with the
partons in the above factorization picture, which is taken as an infrared
cutoff.

Each type of these important contributions gives a large single
logarithm. They may overlap to give a double (leading) logarithm
in some cases. These large logarithms, appearing in a
product with $\alpha_s$, must be organized in order not
to spoil the perturbative expansion. It is known that single logarithms
can be summed to all orders using renormalization group methods, and
double logarithms must be organized by
the technique developed in \cite{CS}.
We shall work in axial gauge $n\cdot A=0$, in which the resummation
technique is developed easily, $n$ being the gauge
vector and $A$ the gauge field.

The diagrams shown in fig.~1 represent
the $O(\alpha_s)$ radiative corrections to the basic factorization
of the pion form factor, which
contain large logarithms mentioned above. In axial
gauge the two-particle reducible diagrams, like figs.~1a and 1b,
have double logarithms from the overlap
of collinear and soft enhancements,
while the two-particle irreducible corrections,
like figs.~1c and 1d, contain only single soft logarithms. This distinction
is consistent with the physical picture: two partons moving in the same
direction can interact with each other through collinear or soft gluons,
while those moving apart from each other can interact only
through soft gluons. Below we shall concentrate on
reducible corrections, and demonstrate how they are summed into
the Sudakov factor.

A careful analysis shows that soft enhancements cancel between
figs.~1a and 1b,
as well as between 1c and 1d, in the asymptotic region with $b\to 0$,
$b$ being the conjugate variable to ${\bf k}_T$.
Therefore, reducible corrections are dominated by
collinear enhancements, and can be absorbed into the pion wave funtion,
which involves similar dynamics.
Irreducible corrections, due to the cancellation of their soft
divergences, are then absorbed into the hard scattering amplitude.
Hence, the factorization picture still holds at least asymptotically
after radiative corrections are included.
The above cancellation of soft divergences is closely related to
the universality of wave functions.
For large $b$, double logarithms are present and the resummation
technique must be implemented.

Based on the above reasoning,
the factorization formula for the pion form factor in the $b$
space is written as \cite{LS},
\begin{eqnarray}
F_{\pi}(Q^{2})&=&\int_0^1 \d x_{1}\d x_{2}\int
\frac{\d^2 {\bf b}}{(2\pi)^2}
\,{\cal P}(x_{2},b,P_{2},\mu)
\nonumber \\
& &\times \,H(x_1,x_2,b,Q,\mu)
\,{\cal P}(x_{1},b,P_{1},\mu)\; ,
\label{3}
\end{eqnarray}
where $\mu$ is the factorization and renormalization scale, and $b$ is
the separation between the two valence quarks. $P_1$ and $P_2$ are momenta
of the incoming and outgoing pions, respectively.
Here we choose the Breit frame, in which $P_1^+=P_2^-=Q/\sqrt{2}$ and
all other components of $P$'s vanish, $Q^2$ being the momentum transfer,
$Q^2=-(P_1-P_2)^2$. Note that eq.~(\ref{3}) depends only on a single $b$,
because the virtual quark line involved in the hard scattering $H$ is thought
of as far from mass shell, and is shrunk to a point \cite{LS}.
The pion wave function ${\cal P}$ includes all leading
logarithmic enhancements at large $b$.
The basic idea of the resummation technique is as follows.
If the double logarithms appear in an exponential form
${\cal P}\sim \exp[-{\rm const.}\times \ln Q\ln (\ln Q/\ln b)]$,
the task will be simplified by studying the derivative of ${\cal P}$,
$\d{\cal P}/\d\ln Q=C{\cal P}$.
It is obvious that the coefficient $C$
contains only large single logarithms, and can be treated by
renormalization group methods. Therefore, working with $C$ one reduces
the double-logarithm problem to a single-logarithm problem.

The two invariants appearing in ${\cal P}$ are $P\cdot n$ and $n^2$,
and by the scale invariance of $n$ in the gluon propagator,
\begin{equation}
N^{\mu\nu}(q)=\frac{-i}{q^2}\left(g^{\mu\nu}-\frac{n^{\mu}q^{\nu}+
q^{\mu}n^{\nu}}{n\cdot q}+n^2\frac{q^{\mu}q^{\nu}}{(n\cdot q)^2}\right)\;,
\end{equation}
${\cal P}$ can only depend on a single large scale
$\nu^2=(P\cdot n)^2/n^2$.
It is then easy to show that the differential operator $\d/\d\ln Q$ can
be replaced by $\d/\d n$:
\begin{equation}
\frac{\d}{\d \ln Q}{\cal P}=-\frac{n^2}{P\cdot n}P^{\alpha}
\frac{\d}{\d n^{\alpha}}{\cal P}\;.
\label{qn}
\end{equation}
The motivation for this replacement is that the momentum $P$ flows through
both quark and gluon lines, but $n$ appears only on gluon lines.
The analysis then becomes easier by studying the $n$, instead
of $P$, dependence.

Applying $\d/\d n_{\alpha}$ to the gluon propagator, we get
\begin{equation}
\frac{\d}{\d n_{\alpha}}N^{\mu\nu}=-\frac{1}{q\cdot n}
(N^{\mu\alpha}q^{\nu}+N^{\nu\alpha}q^{\mu})\;.
\label{dn}
\end{equation}
The momentum $q$ that appears at both ends of a gluon line is contracted
with the vertex, where the gluon attaches. After adding together all
diagrams with different differentiated gluon lines and
using the Ward identity, we arrive at the differential equation of
${\cal P}$ as shown in fig. 2a, in which the square vertex represents
\begin{eqnarray*}
gT^a\frac{n^2}{P\cdot n q\cdot n}P_{\alpha}\;,
\end{eqnarray*}
$T^a$ being the color matrix.
An important feature of the square vertex is that the gluon momentum $q$
does not lead to collinear enhancements because of the nonvanishing
$n^2$. The leading regions of $q$ are then soft and ultraviolet,
in which fig.~2a can be factorized according to fig.~2b at lowest order.
The part on the left-hand side of the dashed line is exactly ${\cal P}$,
and that on the right-hand side is assigned to the coefficient $C$
introduced before.

Therefore, we need a function ${\cal K}$ to organize the soft
enhancements from the first two diagrams in fig.~2b, and ${\cal G}$
for the ultraviolet divergences from the other two diagrams. The soft
substraction employed in ${\cal G}$ is to avoid double counting.
Generalizing the above two functions to all orders,
we derive the differential equation of ${\cal P}$,
\begin{eqnarray}
\frac{\d}{\d \ln Q}{\cal P}(x,b,P,\mu)&=&\left[\,2{\cal K}
(b\mu)+\frac{1}{2}\;{\cal G}(x\nu/\mu)+\frac{1}{2}\;{\cal G}(
(1-x)\nu/\mu)\right]\nonumber \\
& &\times{\cal P}(x,b,P,\mu)\; .
\label{qp}
\end{eqnarray}
The functions ${\cal K}$ and ${\cal G}$ have been calculated
to one loop, and the single logarithms have been organized to give
their evolutions in $b$ and $Q$, respectively \cite{BS}. They possess
individual ultraviolet poles, but their sum ${\cal K}+{\cal G}/2$ is finite
such that Sudakov logarithms are renormalization-group invariant.

Substituting the expressions for ${\cal K}$ and ${\cal G}$ into
eq.~(\ref{qp}), we obtain the solution
\begin{equation}
{\cal P}(x,b,P,\mu)=\exp\left[-\sum_{\xi=x,\;1-x}\;s(\xi,b,Q)\right]{\cal P}
(x,b,\mu)\; .
\label{sp}
\end{equation}
The exponent $s$, grouping the double logarithms in ${\cal P}$,
is expressed in terms of the variables
\begin{eqnarray}
{\hat q} &\equiv & {\rm ln}\left [\xi Q/(\sqrt{2}\Lambda)\right ]
\nonumber \\
{\hat b} &\equiv & {\rm ln}(1/b\Lambda)
\label{11}
\end{eqnarray}
as \cite{LS}
\begin{eqnarray}
s(\xi,b,Q)&=&\frac{A^{(1)}}{2\beta_{1}}\hat{q}\ln\left(\frac{\hat{q}}
{\hat{b}}\right)+
\frac{A^{(2)}}{4\beta_{1}^{2}}\left(\frac{\hat{q}}{\hat{b}}-1\right)-
\frac{A^{(1)}}{2\beta_{1}}\left(\hat{q}-\hat{b}\right)
\nonumber \\
&-&\frac{A^{(1)}\beta_{2}}{16\beta_{1}^{3}}\hat{q}\left[\frac{\ln(2\hat{b})+1}
{\hat{b}}-\frac{\ln(2\hat{q})+1}{\hat{q}}\right]
\nonumber \\
&-&\left[\frac{A^{(2)}}{4\beta_{1}^{2}}-\frac{A^{(1)}}{4\beta_{1}}
\ln\left(\frac{e^{2\gamma-1}}{2}\right)\right]
\ln\left(\frac{\hat{q}}{\hat{b}}\right)
\nonumber \\
&-&\frac{A^{(1)}\beta_{2}}{32\beta_{1}^{3}}\left[
\ln^{2}(2\hat{q})-\ln^{2}(2\hat{b})\right]
\label{sss}
\end{eqnarray}
with $\Lambda\equiv \Lambda_{QCD}$.
The coefficients $A^{(i)}$ and $\beta_{i}$ are
\begin{eqnarray}
& &\beta_{1}=\frac{33-2n_{f}}{12}\;,\;\;\;\beta_{2}=\frac{153-19n_{f}}{24}\; ,
\nonumber \\
& &A^{(1)}=\frac{4}{3}\;,
\;\;\; A^{(2)}=\frac{67}{9}-\frac{\pi^{2}}{3}-\frac{10}{27}n_
{f}+\frac{8}{3}\beta_{1}\ln\left(\frac{e^{\gamma}}{2}\right)\; ,
\label{12}
\end{eqnarray}
where $n_{f}=3$ in this case is the number of quark flavors and
$\gamma$ is the Euler constant. To derive eq.~(\ref{sss}), the choice
of the gauge vector $n^{\mu}\propto (P_1+P_2)^{\mu}$ has been made.

The function ${\cal P}(x,b,\mu)$ and $H$ still contain single logarithms
from ultraviolet divergences, which need to be
summed using their renormalization group equations \cite{BS}:
\begin{eqnarray}
& &{\cal D}{\cal P}(x,b,\mu)=-2\,\gamma_q\,{\cal P}(x,b,\mu)
\nonumber \\
& &{\cal D}\,H(x_i,b,Q,\mu)=4\,\gamma_q\,H(x_i,b, Q,\mu)\; ,
\label{re}
\end{eqnarray}
with
\begin{eqnarray}
{\cal D}=\mu\frac{\partial}{\partial \mu}+\beta(g)\frac{\partial}
{\partial g}\;.
\end{eqnarray}
$\gamma_q=-\alpha_s/\pi$ is the quark anomalous dimension
in axial gauge. Solving eq.~(\ref{re}),
the large-$b$ asymptotic behavior of $\cal P$
can be summarized as
\begin{eqnarray}
{\cal P}(x,b,P,\mu)&=&\exp\left[-\sum_{\xi=x,\;1-x}\;s(\xi,b,Q)
-2\int_{1/b}^{\mu} \frac{d\bar{\mu}}{\bar{\mu}}\gamma
_q(g(\bar{\mu}))\right]
\nonumber \\
& &\times {\cal P}(x,b,1/b)\;,
\end{eqnarray}
where
\begin{eqnarray}
{\cal P}(x,b,1/b)=\phi (x,b)+O(\alpha_{s}(1/b))\; .
\label{10}
\end{eqnarray}
The evolution of $\cal P$ in $b$, denoted by $O(\alpha_s)$,
has been neglected. The $b$ dependence in $\phi$, corresponding to the
intrinsic transverse momentum dependence of
the pion wave function \cite{BSW,JK}, will not be considered here.

Similarly, the renormalization group analysis applied to $H$ gives
\begin{eqnarray}
H(x_i,b,Q,\mu)
&=&\exp\left[-4\,\int_{\mu}^{t}\frac{\d\bar{\mu}}{\bar{\mu}}
\,\gamma_q(g(\bar{\mu}))\right]\nonumber \\
& &\times H(x_i,b,Q,t)\;,
\label{13}
\end{eqnarray}
where $t$ is the largest mass scale involved in the hard scattering,
\begin{equation}
t=\max(\sqrt{x_{1}x_{2}}Q,1/b)\; .
\label{9}
\end{equation}
The scale $\sqrt{x_1x_2}Q$ is associated with the
longitudinal momentum of the
hard gluon and $1/b$ with its transverse momentum.
Combining all the exponents derived above, we
obtain the lowest-order expression for the pion form factor,
\begin{eqnarray}
F_{\pi}(Q^2)&=& 16\pi{\cal C}_F\int_{0}^{1}\d x_{1}\d x_{2}\,
\phi(x_{1})\phi(x_{2})
\int_{0}^{\infty} b\d b\, \alpha_{s}(t)\, K_{0}(\sqrt{x_{1}x_{2}}Qb)
\nonumber \\
& & \times \exp[-S(x_{1},x_{2},b,Q)]\; ,
\label{15}
\end{eqnarray}
where the complete Sudakov logarithms are given by
\begin{equation}
S(x_{1},x_{2},b,Q)=\sum_{i=1}^{2}\left[s(x_{i},b,Q)+s(1-x_{i},b,Q)\right]-
\frac{2}{\beta_{1}}{\rm ln}\frac{\hat{t}}{\hat{b}}\; ,
\label{16}
\end{equation}
with $\hat{t}={\rm ln}(t/\Lambda)$.
${\cal C}_F$ is the color factor defined by ${\rm tr}(T^aT^a)=N_c
{\cal C}_F$, $N_c$ being the number of colors.
$K_{0}$ is the modified Bessel function of order
zero, which is the Fourier transform of the gluon propagator
to the $b$ space.

Variation of $S$ with $b$ has been displayed in \cite{LS}, which shows
a strong falloff in the large $b$ region, and vanishes for $b > 1/\Lambda$.
Hence, Sudakov suppression
selects components of the pion wave functions with
small spatial extent $b$, and makes the hard scattering more perturbative.
If $b$ is small, $\alpha_s$ with its argument set to $t$ as in
eq.~(\ref{9}) will be small,
regardless of the values of the $x$'s.
When $b$ is large and $x_1x_2Q^2$ is small, $\alpha_s$
is still large. However,
the Sudakov factor in eq.~(\ref{15}) strongly suppresses this region.
Since the main contributions to the factorization formula
come from the small $b$, or short-distance, region,
the perturbation theory becomes relatively self-consistent.
It is also easy to show that the modified factorization formula including
Sudakov corrections reduces to the standard one as
$b\to 0$, where the important logarithms diminish.

\vskip 2.0cm

\cl{\large \bf 3. The Decay $B\to\pi l \nu$}
\vskip 0.5cm
We now generalize the above analysis of the pion form factor to
the exclusive processes involving both light and heavy
mesons. In particular, we work with the semi-leptonic decay
$B\to\pi l \nu$. PQCD is appropriate to this process when the pion is
energetic, because the $b$ quark mass provides a large scale.
We shall analyze the leading regions of radiative corrections to
such a process, and derive the Sudakov factor including both
leading and next-to-leading logarithms.
The amplitude of this decay is written as
\begin{equation}
A(P_1,P_2)=\frac{G_F}{\sqrt{2}}V_{ub}{\bar \nu}\gamma_{\mu}(1-\gamma_5)l
\langle\pi (P_2)|{\bar b}\gamma^{\mu}u|B(P_1)\rangle\;,
\end{equation}
where the four-fermion interaction with the Fermi coupling constant
$G_F$ has been inserted. $P_1$ and $P_2$ are momenta of the $B$ meson
and the pion, respectively.
We start with the lowest-order factorization for the matrix
element $M^{\mu}=
\langle\pi (P_2)|{\bar b}\gamma^{\mu}u|B(P_1)\rangle$
with an exchanged hard gluon as shown in
fig.~3, the left-hand side being the $B$ meson at rest and the
right-hand side a fast-recoiling pion. The heavy $b$ quark is represented
by a bold line. The symbol
$\times$ denotes the electroweak vertex
with the KM matrix element $V_{ub}$, from which a lepton pair emerges.

Parton momenta are assigned as in fig.~3. The $b$ quark carries
$P_1-k_1$,
and the accompanying light quark carries $k_1$. They satisfy the
on-shell conditions $(P_1-k_1)^2\approx m_b^2$, $P_1^2=m_B^2$ and
$k_1^2\approx 0$, $m_B$ being the $B$ meson mass. In the Breit frame
$P_1$ has the components $P_1^+=P_1^-=m_B/\sqrt{2}$ and
vanishing transverse components.
$k_1$ may have a large minus component with small amount of
transverse components ${\bf k}_{1T}$, which will serve
as the infrared cutoff
of Sudakov corrections below.
The assignment of parton momenta on the pion side is similar to that
in the case of the pion form factor as shown in fig.~3.
The large component of $P_2$ is
$P_2^+=\eta m_B/\sqrt{2}$, $\eta$ being related to
the energy fraction of the pion by $P_2^0=\eta m/2$.  The physical range of
$\eta$ is $0 \le \eta \le 1$, since the pion carries away at most
half of the rest energy of the $B$ meson.
The transverse momentum associated with the valence quarks
of the pion is denoted by ${\bf k}_{2T}$.
The invariant mass of the lepton-neutrino pair
produced in this decay
is given by $m_l^2=(P_1-P_2)^2=(1-\eta)m_B^2$.

We then consider radiative corrections to the basic factorization for
the heavy-to-light transition.
The essential step is again to find out the leading regions of radiative
corrections in axial gauge. For reducible corrections on
the pion side, the conclusion is the same as before: they produce double
logarithms with soft ones cancelled in the asymptotic region, and
can be absorbed into the pion wave function, which give rise to the evolution
of the wave function.
Irreducible corrections, with an extra gluon
connecting a quark in the pion and a quark in the $B$ meson,
give only soft divergences, which also cancel
asymptotically. Hence, they are absorbed
into the hard scattering amplitude.

On the left-hand
side, three diagrams showing the $O(\alpha_s)$
reducible radiative corrections, are exhibited in fig.~4.
Fig. 4a, giving the self-energy correction to the massive $b$ quark,
produces only soft enhancements, and is thus not leading.
If $k_1^-$ is small, collinear divergences in figs. 4b and 4c,
which arise from the loop momentum with a large component parallel to $k_1$,
will not be pinched, and they also give only soft enhancements.
This is consistent with the physical picture that the
soft light quark can not interact with the heavy quark through a fast
moving gluon. However, from the $B$ meson wave functions given in
section 4, it is observed that there is
substantial probability of finding the light quark with $k_1^-$ of order
$m_b$, even though the wave functions peak at small $k_1^-$.
Therefore, figs.~4b and 4c contribute collinear enhancements.
Note that fig. 4b gives soft divergences which are not completely cancelled
by those from figs.~4a and 4c even in the asymptotic region.
In conclusion, figs.~4b and 4c indeed contain double logarithms,
which must be organized by the resummation technique.

Since the collinear enhancements on the $B$ meson side
are less important due to the suppression of the wave functions, reducible
corrections are basically dominated by soft enhancements,
and can be absorbed into the $B$
meson wave function, which is also dominated by
soft dynamics. This absorption of reducible
corrections is similar to that on the pion side,
where reducible corrections are dominated by collinear enhancements.
Therefore, the factorization picture in fig.~3 still holds after radiative
corrections are included.

With the above reasoning, we can write down the factorization formula
for the decay $B\to\pi l\nu$ in the transverse
configuration space,
\begin{eqnarray}
M^\mu&=&\int_0^1 \d x_{1}\d x_{2}\int
\frac{\d^2 {\bf b}_1}{(2\pi)^{2}}\frac{\d^2 {\bf b}_2}{(2\pi)^{2}}
\,{\cal P}_\pi(x_{2},{\bf b}_2,P_{2},\mu)
\nonumber \\
& &\times \,{\tilde H}^\mu(x_1,x_2,{\bf b}_1,{\bf b}_2,m,\mu)
\,{\cal P}_B(x_{1},{\bf b}_1,P_{1},\mu)\; ,
\label{fbd}
\end{eqnarray}
where both the pion and $B$ meson wave functions,
${\cal P}_\pi$ and ${\cal P}_B$, contain leading double
logarithms. Here we have to introduce two $b$'s, because the virtual
quark line in the hard scattering may not be far from mass shell, and can not
be shrunk to a point. This is contrary to the case of the pion
form factor, and detailed explanation will be given later.
Therefore, we need $b_1$ to denote the separation between the two valence
quarks of the $B$ meson, and $b_2$ for the pion.
The approximation $m_b\approx m_B =m$ has been made.
The momentum fraction $x_1$ is defined as $k_1^-/P_1^-$. ${\tilde H}^{\mu}$
is the Fourier transform of the hard scattering amplitude derived from
fig.~3, whose explicit expression will be given in section 4.
The resummation of the double logarithms in ${\cal P}_\pi$
has been performed in the previous section. Here we quote the results directly
with $Q$ set to $\eta m$ and $n_f$ set to 4 \cite{YS}.
In the below we shall concentrate on ${\cal P}_B$.

There are two major difficulities in summing up the double logarithms in
figs.~4b and 4c. Firstly, fig.~4a, which gives only
single logarithms, must be excluded. Secondly, there are many invariants
that can be constructed from $P_1$, $k_1$ and $n$
such as $P_1^2$, $P_1\cdot k_1$, $P_1\cdot n$,
$k_1\cdot n$ and $n^2$,
which are involved in ${\cal P}_B$.
In the pion case the number of invariants is much less. There are
only $P\cdot n$ and $n^2$, and the resummation is thus simpler.
The fact that ${\cal P}_B$ contains many invariants fails the technique
of replacing $\d/\d m$ by $\d/\d n$
as introduced in section 3, because in this case some large scales like
$P_1^2$ can not be related to $n$.

However, the above difficulities can be overcomed by applying
the eikonal approximation to the heavy quark line as shown in fig. 5.
In the collinear region with the loop momentum parallel to $k_1$
and in the soft region, the $b$ quark line can be
replaced by an eikonal line:
\begin{equation}
\frac{(\not P_1-\not k_1+\not q+m)\gamma^{\alpha}}{(P_1-k_1+q)^2-m^2}
\approx \frac{(P_1-k_1)^{\alpha}}{(P_1-k_1)\cdot q}+{\rm R}\;,
\label{eik}
\end{equation}
where the remaining part ${\rm R}$ either vanishes as contracted with
the matrix structure of the $B$ meson wave function, or is less leading.
The factor $1/[(P_1-k_1)\cdot q]$ is associated with the eikonal
propagator, and the numerator $(P_1-k_1)^{\alpha}$ is absorbed
into the vertex,
where a gluon attaches to the eikonal line.
The physics involved in this approximation
is that a soft gluon or a gluon moving parallel to $k_1$ can not explore
the details of the $b$ quark, and its dynamics can be factorized.
The idea is similar to that employed in HQET.
An explicit evaluation of radiative corrections confirms this approximation.
By this means, the first difficulty is resolved, because self-energy
diagrams of an eikonal line are excluded by definition \cite{C}.

The eikonal approximation also reduces the number
of large invariants involved in ${\cal P}_B$.
We have the scale invariance in
$P_1-k_1\approx P_1$ as shown by Feynman rules for an eikonal line
in eq.~(\ref{eik}), which corresponds to the flavor symmetry in HQET,
in addition to the scale invariance in $n$.
Hence, $P_1$ does not lead to
a large scale, and the only remaining large scale is $k_1^-$,
which must appear in the ratios
$(k_1\cdot n)^2/n^2$ and $(k_1\cdot P_1)^2/P_1^2$.
An explicit lowest-order investigation shows that the second scale
$(k_1\cdot P_1)^2/P_1^2$ in fact does not exist.
Therefore, with the eikonal
approximation the problem is simplified to the one
in analogy with the light-meson case. Now ${\cal P}_B$ depends only
on the single large scale $\nu'^2=(k_1\cdot n)^2/n^2$, and
$\d/\d k_1^-$ can be replaced by $\d/\d n$.

Following the same procedures as in section 2,
the differential equation of ${\cal P}_B$ is derived as
\begin{equation}
\frac{\d}{\d \ln m}{\cal P}_B=\frac{\d}{\d \ln k_1^-}{\cal P}_B=
\left[{\cal K}(b\mu)+\frac{1}{2}{\cal G}(\nu'/\mu)\right]{\cal P}_B\;,
\label{dfb}
\end{equation}
where the lowest-order ${\cal K}$ can be obtained
from fig.~6a, and ${\cal G}$
from fig.~6b, with the square vertex representing
\begin{eqnarray*}
gT^a\frac{n^2}{k_1\cdot n q\cdot n}k_1^{\alpha}\;.
\end{eqnarray*}
Note the absence of the diagrams corresponding to self-energy corrections
to the eikonal line.

Comparing fig.~6a with 2a, we find that the evaluation of ${\cal K}$
for the $B$ meson is similar to that for the pion except the third
diagram. This extra diagram is finite without ultraviolet and infrared
divergences, as justified by its integral
\begin{equation}
g^2{\cal C}_F
\int\frac{\d^4q}{(4\pi)^4}\frac{n^2k_1^{\alpha}
P_1^{\beta}}{k_1\cdot n q\cdot n P_1\cdot q}\frac{iN_{\alpha\beta}}{q^2}
e^{i{\bf q}_T\cdot {\bf b}}\;.
\label{kex}
\end{equation}
Hence, it does not spoil the renormalization-group invariance
of the Sudakov logarithms. For the
specific choice $n\propto P_1$ as in the case of the pion form factor,
it is easy to show that (\ref{kex}) vanishes.
Therefore, the functions ${\cal K}$ and ${\cal G}$
for the $B$ meson are in fact the same as those for the pion.

Substituting the expressions of ${\cal K}$ and ${\cal G}$
into eq.~(\ref{dfb}), we obtain the solution
\begin{equation}
{\cal P}_B(x_1,b_1,P_1,\mu)=\exp\left[-s(x_1,b_1,m)\right]{\cal P}_B
(x_1,b_1,\mu)\; ,
\label{sb}
\end{equation}
where the exponent $s$ is given by eq.~(\ref{sss}) but with
$n_f=4$ \cite{YS}.
Summing up single logarithms in ${\cal P}_B(x_1, b_1, \mu)$, eq.~(\ref{sb})
becomes
\begin{eqnarray}
{\cal P}_B(x_1,b_1,P_1,\mu)&=&\exp\left[-s(x_1,b_1,m)
-2\int_{1/b_1}^{\mu} \frac{d\bar{\mu}}{\bar{\mu}}\gamma
_q(g(\bar{\mu}))\right]
\nonumber \\
& &\times \phi_B(x_1,b_1)+O(\alpha_{s}(1/b_1))\;,
\label{sop}
\end{eqnarray}
where the anomalous dimension $\gamma_q$ is the same as before.
Again, the evolution of ${\cal P}_B$ in $b_1$
and the intrinsic $b_1$ dependence of the wave function $\phi_B$
will be neglected below.
Including the summation of single logarithms in ${\tilde H}^\mu$
and the results from ${\cal P}_\pi$, we obtain the complete Sudakov exponent
\begin{eqnarray}
S(x_i,b_i,\eta,m)&=&s(x_1,b_1,m)+
s(x_2,b_2,\eta m)+s(1-x_2,b_2,\eta m)
\nonumber \\
& &-\frac{1}{\beta_{1}}\left({\rm ln}\frac{\hat{t}}{\hat{b_1}}
+{\rm ln}\frac{\hat{t}}{\hat{b_2}}\right)\;.
\label{sda}
\end{eqnarray}
The variable ${\hat b}_i$ is defined as in eq.~(\ref{11}), and
$t$ is the largest scale involved in the hard scattering, which will be
discussed in section 4.
Similar expression to eq.~(\ref{sda}) for the pion form factor,
which also involves two $b$'s, has been obtained in \cite{L}.
\vskip 2.0cm

\centerline{\large \bf 4. Numerical Results}
\vskip 0.5cm

Having derived the Sudakov factor for the process $B\to\pi l\nu$,
we now evaluate its decay rate, and examine how much contribution comes
from the perturbative region with small $b_i$.
The explicit formula for lowest-order $H^{\mu (a)}$ from fig. 3a
is written as
\begin{eqnarray}
H^{\mu (a)}&=&
{\rm tr}\left[\gamma_\alpha\frac{\gamma_5 \not P_2}{\sqrt{2N_c}}\gamma^\mu
\frac{\not P_1-x_2\not P_2 +\not{\bf k}_{2T}+m}
{(P_1-x_2P_2+{\bf k}_{2T})^2-m^2}\gamma^{\alpha}
\frac{(\not P_1+m)\gamma_5}{\sqrt{2N_c}}\right]
\nonumber \\
& &\times \frac{-g^2N_c{\cal C}_F}{(x_2P_2-k_1+{\bf k}_{2T})^2}
\nonumber \\
&=&\frac{4(1+x_2\eta)g^2{\cal C}_Fm^2}{[x_2\eta m^2+{\bf k}_{2T}^2]
[x_1x_2\eta m^2+({\bf k}_{1T}-{\bf k}_{2T})^2]}P_2^\mu\;,
\label{ha}
\end{eqnarray}
where the factors $\gamma_5 \not P_2/\sqrt{2N_c}$ and
$(\not P_1+m)\gamma_5/\sqrt{2N_c}$ come from the matrix structures of
the pion and $B$ meson wave functions, respectively \cite{YS}.
The relation $k_1^-=x_1m/\sqrt{2}$ has been inserted. In the second
expression the ${\bf k}_{2T}$ dependence in the fermion
propagator is not neglected. For the pion form factor the corresponding
transverse momentum dependence
is negligible, because there is a factor $x_2$
in the numerator, which cancels the singularity from $x_2\to 0$
in the denominator. However, for the $B$ meson decays, due to the massiveness
of the $b$ quark, such a cancellation does not appear as shown in
eq.~(\ref{ha}). To ensure that the virtual quark be part of the hard
scattering, ${\bf k}_{2T}$ must be retained.
Therefore, with the inclusion of transverse momenta, we need not to
perform the substraction of an on-shell fermion propagator
from the hard scattering as in \cite{YS}.

Similarly, the expression for lowest-order $H^{\mu(b)}$ is given by
\begin{eqnarray}
H^{\mu (b)}&=&
{\rm tr}\left[\gamma_\alpha\frac{\gamma_5 \not P_2}{\sqrt{2N_c}}
\gamma^\alpha\frac{\not P_2 +\not k_1}{(P_2+k_1)^2}\gamma^{\mu}
\frac{(\not P_1+m)\gamma_5}{\sqrt{2N_c}}\right]
\nonumber \\
& &\times \frac{-g^2N_c{\cal C}_F}{(x_2P_2-k_1+{\bf k}_{2T})^2}
\nonumber \\
&=&\frac{4g^2{\cal C}_Fx_1\eta m^2}{[x_1\eta m^2+{\bf k}_{1T}^2]
[x_1x_2\eta m^2+({\bf k}_{1T}-{\bf k}_{2T})^2]}P_1^\mu
\nonumber \\
& &-\frac{4g^2{\cal C}_Fx_1 m^2}{[x_1\eta m^2+{\bf k}_{1T}^2]
[x_1x_2\eta m^2+({\bf k}_{1T}-{\bf k}_{2T})^2]}P_2^\mu\;.
\label{hb}
\end{eqnarray}
To derive the second expression, we have replaced $k_1^\mu$ by
\begin{equation}
\frac{P_2\cdot k_1}{P_1\cdot P_2}P_1^\mu+
\left(\frac{P_1\cdot k_1}{P_1\cdot P_2}-
\frac{2}{\eta}\frac{P_2\cdot k_1}{P_1\cdot P_2}\right)P_2^\mu\;.
\end{equation}
Here the transverse momentum ${\bf k}_{1T}$ in the fermion propagator
is negligible, because the singularity from $x_1\to 0$ is removed by
the numerator. However, we will keep it for consistency.

Performing the Fourier transform of
eqs.~(\ref{ha}) and (\ref{hb}), the matrix element $M^\mu$
can be written as
\begin{equation}
M^\mu=f_1P_1^\mu+f_2P_2^\mu\;,
\end{equation}
in which the factorization formulas for the form factors $f_1$ and
$f_2$ are given by
\begin{eqnarray}
f_1&=& 16\pi{\cal C}_Fm^2\int_{0}^{1}\d x_{1}\d x_{2}\,
\int_{0}^{\infty} b_1\d b_1 b_2\d b_2\,
\phi_B(x_{1})\phi_\pi(x_{2})
\nonumber \\
& &\times x_1\eta h(x_1,x_2,b_1,b_2,\eta,m)
\exp[-S(x_i,b_i,\eta,m)]\;,
\label{f1}
\end{eqnarray}
and
\begin{eqnarray}
f_2&=& 16\pi{\cal C}_Fm^2\int_{0}^{1}\d x_{1}\d x_{2}\,
\int_{0}^{\infty} b_1\d b_1 b_2\d b_2\,
\phi_B(x_{1})\phi_\pi(x_{2})
\nonumber \\
& &\times [-x_1h(x_1,x_2,b_1,b_2,\eta,m)+(1+x_2\eta)
h(x_2,x_1,b_2,b_1,\eta,m)]
\nonumber \\
& &\times \exp[-S(x_i,b_i,\eta,m)]\;,
\label{f2}
\end{eqnarray}
respectively, with
\begin{eqnarray}
h(x_1,x_2,b_1,b_2,\eta,m)&=&
\alpha_{s}(t)K_{0}(\sqrt{x_{1}x_{2}\eta}mb_2)
\nonumber \\
& &\times [\theta(b_1-b_2)K_0(\sqrt{x_1\eta}mb_1)I_0(\sqrt{x_1\eta}mb_2)
\nonumber \\
& &\;\;\;\;
+\theta(b_2-b_1)K_0(\sqrt{x_1\eta}mb_2)I_0(\sqrt{x_1\eta}mb_1)]\;.
\label{dh}
\end{eqnarray}
$I_0$ is the modified Bessel function of order zero.
{}From eqs.~(\ref{ha}) and (\ref{hb}),
we choose the largest scale $t$ associated with the
hard gluon as
\begin{equation}
t=\max(\sqrt{x_1x_2\eta}m,1/b_1,1/b_2)\;.
\end{equation}
Basically, the above expressions for $f_i$ are similar to that for
the pion form factor but with different Sudakov logarithms.

To evaluate $f_i$, we consider the following two models of $\phi_B(x)$,
which have been adopted in \cite{YS}.
They are the oscillator wave function \cite{BW}
\begin{eqnarray}
\phi_B^{(I)}(x)&=&N\sqrt{x(1-x)}\exp\left[-\frac{m_B^2}
{2\omega^2}\left(\frac{1}{2}-x-\frac{m_b^2}{2m_B^2}\right)^2\right]
\nonumber \\
&\approx& N\sqrt{x(1-x)}\exp\left[-\frac{m^2}
{2\omega^2}x^2\right]\;,
\end{eqnarray}
in our approximation $m_b\approx m_B$, and \cite{S}
\begin{equation}
\Phi_B^{(II)}(x,{\bf k}_T)=N'\left[C+\frac{m_b^2}{1-x}+\frac{{\bf k}_T^2}
{x(1-x)}\right]^{-2}\;,
\end{equation}
where $x$ is the momentum fraction of the light quark in the
$B$ meson.
The parameters in $\phi_B^{(I)}$ are $\omega=0.4$ GeV and $m=5.28$ GeV.
The constant $N$ is determined by the normalization of the wave
function:
\begin{equation}
\int_0^1\d x\phi_B^{(I)}(x)=\frac{f_B}{2\sqrt{3}} \;,
\end{equation}
$f_B=160$ MeV being the $B$ meson decay constant \cite{BLS},
which leads to $N=2.24$ GeV.

Similarly, the constants $N'$ and $C$ in
$\Phi_B^{(II)}$ are determined by the following normalizations \cite{S}:
\begin{eqnarray}
& &\int_0^1\d x\int
\d^2 {\bf k}_T\Phi_B^{(II)}(x,{\bf k}_T)
=\frac{f_B}{2\sqrt{3}} \;,
\nonumber \\
& &\int_0^1\d x\int
\d^2 {\bf k}_T[\Phi_B^{(II)}(x,{\bf k}_T)]^2
=\frac{1}{2}\;,
\end{eqnarray}
from which $N'=1.232$ GeV$^3$ and $C=-0.99993m^2$ are obtained.
The Fourier transform of $\Phi_B^{(II)}$ gives
\begin{eqnarray}
\phi_B^{(II)}(x,b)&=&\int\d^2 {\bf k}_T
\Phi_B^{(II)}(x,{\bf k}_T)e^{i{\bf k}_T\cdot {\bf b}}
\nonumber \\
&=&\frac{\pi N'bx^2(1-x)^2}{\sqrt{m^2x+Cx(1-x)}}
K_1\left(\sqrt{m^2x+Cx(1-x)}b\right)\;,
\end{eqnarray}
with $K_1$ the modified Bessel function of order one. As stated before,
we neglect the intrinsic $b$ dependence of the wave function, and
define
\begin{equation}
\phi_B^{(II)}(x)=\lim_{b\to 0}\phi_B^{(II)}(x,b)=
\frac{\pi N'x(1-x)^2}{m^2+C(1-x)}\;.
\end{equation}
Obviously, both of the models peak at small $x$, which signifies
the soft dynamics in the $B$ meson. However, the probability
at intermediate $x$ is indeed comparable at least in model II,
and the resummation of double logarithms performed in section 3
is essential. At last, we adopt the Chernyak-Zhitnitsky model
\cite{CZ} for the pion wave function:
\begin{equation}
\phi_\pi(x)=5\sqrt{3}f_\pi x(1-x)(1-2x)^2
\end{equation}
with $f_\pi=93$ MeV the pion decay constant.

We are now ready to evaluate $f_1$ and $f_2$ numerically
in eqs.~(\ref{f1}) and (\ref{f2}). Results of $f_i$ for the two
models of $\phi_B$, with $b_1$ and $b_2$
integrated up to the same cutoff $b_{1c}=b_{2c}=b_c$, are shown
in fig.~7. It is found that at $\eta=0.3$
approximately 50\% of the contribution
to $f_i$ comes from the region where $\alpha_s(1/b_c)< 1$, or equivalently,
$b_c< 0.5\Lambda$.
At $\eta=0.4$, 55\% of the contribution is accumulated in this
perturbative region. As $\eta=1$, perturbative contribution
has reached 70\%.
It implies that the PQCD analysis of the decay $B\to\pi l \nu$
in the range of $\eta > 0.3$
is relatively reliable according to the criteria given in \cite{LS,L}.
Therefore, we shall
accept the modified PQCD predictions for $\eta \ge 0.3$, and
use them to evaluate the differential decay rate.
It is also observed that the expressions with $\phi_B^{(II)}$ employed
are more perturbative, because the large $k_1^-$ region is less
suppressed by the wave function, and the double logarithms are stronger.
The predictions from $\phi_B^{(I)}$ are larger, because
$\phi_B^{(I)}$ enhances the contribution from the end-point region of $x_1$.

In the approach of \cite{YS} where the transverse
momentum dependence was not included,
instead, energies of the virtual quark and gluon
involved in the hard scattering
were taken as the ultraviolate and infrared cutoffs
of radiative corrections,
respectively. The resulting Sudakov logarithm proportional to $\ln m/k_1^-$,
however, gives weaker suppression. From the steepest descent approximation
of their Sudakov factor, the saddle point at $k_1^-$ around $1.4\Lambda$
for $\eta=1$ has been found. Using the same method, we determine the
saddle point of our Sudakov factor in eq.~(\ref{sda}) by
\begin{equation}
\frac{\partial S}{\partial b_1}=\frac{\partial S}{\partial b_2}=0\;,
\end{equation}
from which a larger scale $1/b_1=10\Lambda$ for $\Lambda=0.1$ GeV, or $1/b_1=
6\Lambda$ for $\Lambda=0.2$ GeV is obtained. It is then obvious that
the perturbative expansion in our approach is more reliable. At this large
scale the double logarithms contained in
the $O(\alpha_s)$ radiative corrections
to the hard scattering with a triple gluon vertex, which have been considered
in \cite{YS}, are in fact not important.

We observe that the magnitude of $f_2$ is much
larger than that of $f_1$, especially in the small $\eta$ region.
This fact is consistent with their behaviors in the soft pion limit
as obtained in the framework of HQET \cite{BLN}, in which $f_2$
is found to have a pole at $\eta\to 0$:
\begin{equation}
\lim_{\eta\to 0}f_2\approx \frac{2f_{B^*}}{\eta f_\pi}g_{BB^*\pi}\;,
\label{sf2}
\end{equation}
for $m_{B^*}\approx m_B$, $m_{B^*}$ being the $B^*$ meson mass.
In the above expression
$g_{BB^*\pi}$ is the $BB^*\pi$ coupling constant, and
$f_{B^*}$ the decay constant of the $B^*$ meson. Assuming
$\eta= 0.3$, $f_{B^*}\approx 1.1 f_B$ \cite{N2} and
$g_{BB^*\pi}\approx 0.75$ \cite{YL} in
eq.~(\ref{sf2}), we obtain $f_2=9.5$ which is close to
our predictions at $\eta=0.3$. Certainly, this extrapolation of the soft
pion theorems may not be reliable, but it is interesting to observe that
it is consistent with the PQCD results at the middle value of $\eta$.
$f_1$ in the soft pion limit vanishes like $1-\sqrt{m_{B^*}/m_B}$ \cite{BLN}.

With results of $f_1$ and $f_2$, we can compute the differential
decay rate of $B^0\to\pi^- l^+\nu$ with vanishing
lepton masses \cite{YS}:
\begin{eqnarray}
\frac{\d \Gamma}{\d \eta}&=&|V_{ub}|^2\frac{G_F^2m^5\eta^3}{768\pi^3}
|f_1+f_2|^2
\nonumber \\
&\equiv &|V_{ub}|^2R(\eta)\;,
\end{eqnarray}
for $\eta>0.3$, where the second formula defines the function $R(\eta)$.
Results of $R(\eta)$ are shown in fig.~8,
from which the decrease of $\d\Gamma/\d
\eta$ with $\eta$ is observed. The behavior predicted here is opposite to
that given in \cite{YS}, which shows an increase in $\eta$ starting
with zero at $\eta=0.5$.
Such a dip at the middle value of $\eta$ is attributed to
the substraction of an on-shell fermion propagator from the hard
scattering \cite{YS},
which is, however, not necessary in our treatment.
Furthermore, the predictions in \cite{YS} for the differential decay rate
are almost 10$^3$ times smaller than ours. The reason is again traced
back to the substraction of the fermion propagator.
Therefore, the transverse momentum dependence plays an essential role
in the analysis of $B$ meson decays.

The differential decay rate in the soft pion limit can be obtained
using eq.~(\ref{sf2}):
\begin{equation}
\lim_{\eta\to 0}R(\eta)=
\frac{G_F^2m^5\eta}{192\pi^3}\frac{f_{B^*}^2}{f_\pi^2}
g_{BB^*\pi}^2\;,
\label{spl}
\end{equation}
which shows a linear relation with $\eta$.
Extrapolating eq.~(\ref{spl}) to intermediate
$\eta$ as shown in fig.~8,
we observe a fair match between the soft pion and
PQCD predictions around $\eta=0.3$. It implies that
the modified PQCD formalism is successful at large $\eta$, but becomes
worse quickly in the soft pion limit. Similarly, the soft pion technique
is appropriate at small $\eta$, but gives an overestimation
to the considered process in the perturbative region.
The overlap of these two approaches indicates the transition
of the $B$ meson decays to PQCD at middle values of $\eta$, and the
complementarity between the soft pion theorems and modified perturbative
formalism \cite{CL,CL2}.

We then estimate the total decay rate $\Gamma$
by integrating $\d\Gamma/\d\eta$ using eq.~(\ref{spl}) for $\eta < 0.3$
and using our PQCD predictions for $\eta > 0.3$.
We obtain $0.6\times 10^{-11} |V_{ub}|^2$ GeV from the soft pion theorems,
and $1.9\times 10^{-11} |V_{ub}|^2$ and
$0.7\times 10^{-11} |V_{ub}|^2$ GeV for the use of $\phi_B^{(I)}$ and
$\phi_B^{(II)}$, respectively, from the modified PQCD formalism. Their
sum gives $\Gamma\approx 2.5\times 10^{-11} |V_{ub}|^2$ GeV for
model I, and $1.3\times 10^{-11} |V_{ub}|^2$ for model II.
They correspond to
branching ratios $0.5\times 10^2|V_{ub}|^2$ and $0.26\times 10^2|V_{ub}|^2$,
respectively, for the total width of
the $B^0$ meson is $(0.51\pm 0.02)\times 10^{-9}$ MeV \cite{RPP}.
Current experimental limit on the branching ratio of $B^0\to
\pi^- l^+\nu$ is $3.3\times 10^{-4}$ \cite{O}. We then extract the
the matrix element $|V_{ub}|< 2.6\times 10^{-3}$
from model I and $|V_{ub}|< 3.6\times 10^{-3}$ from model II.
These upper limits are close to the value 0.003
given in the literature \cite{RPP}.

\vskip 2.0cm

\centerline{\large \bf 5. Conclusion}
\vskip 0.5cm

In this paper we have applied the resummation technique to the
semi-leptonic decay $B\to\pi l\nu$, and derived the Sudakov factor
up to next-to-leading logarithms in this heavy-to-light process.
The idea is to apply the eikonal approximation  to the heavy $b$
quark line such that its nonleading self-energy diagram is excluded, and
the number of large scales involved in the $B$ wave function is reduced.
The resummation of double logarithms in the heavy meson is then simplified
to the one in analogy with the light meson.
The modified PQCD calculation of the differential decay rate
including Sudakov effects has been examined and found to be
reliable for $\eta$ above 0.3.
By combining our predictions with the soft pion results and
comparing them with experimental data, we
have estimated the total decay rate, from which the upper limit 0.003
for the KM matrix element $|V_{ub}|$ is obtained.

We do not observe the dip at $\eta=0.5$ for the differential decay
rate as predicted in \cite{YS}, which arises from the substraction
of an on-shell fermion propagator from the hard scattering. This
substraction is not necessary in our analysis
because of the inclusion the transverse
momentum dependence. The behavior in $\eta$ is similar to that
of the pion form factor in $Q^2$, which
is opposite to the results of \cite{YS}. We do not consider
the evolution in $b$ and the intrinsic transverse momentum dependence
of wave functions in this work. However, we estimate that these two
modifications cancel at least partially, since the former gives
an enhancement \cite{L} to the decay rate,
but the latter leads to a suppression \cite{JK}.
Certainly, this subject needs more detailed investigation.
Our formalism can be easily applied to a similar semi-leptonic decay
$B\to \rho l\nu$ \cite{YS}
and other non-leptonic $B$ meson decays,
which will be published elsewhere.

\vskip 0.5cm
We thank G.L. Lin, M. Neubert, G. Sterman and Y.P. Yao for helpful discussions.
This work was supported by the National
Science Council of R.O.C. under Grant Nos. NSC83-0208-M194-019T and
NSC83-0208-M001-015.

\newpage
\cl{\large \bf Figure Captions}
\vskip 0.5cm

\noindent
{\bf Fig. 1.} $O(\alpha_s)$ radiative
corrections to the basic factorization of the pion form factor.
\vskip 0.5cm

\noindent
{\bf Fig. 2.} Graphic representation of eq.~(\ref{qp}).
\vskip 0.5cm

\noindent
{\bf Fig. 3.} Lowest-order factorization for the decay $B\to\pi l\nu$.
\vskip 0.5cm

\noindent
{\bf Fig. 4.} $O(\alpha_s)$ radiative
corrections to the $B$ meson wave function.
\vskip 0.5cm

\noindent
{\bf Fig. 5.} Eikonal approximation for the $b$ quark line.
\vskip 0.5cm

\noindent
{\bf Fig. 6.} Lowest-order diagrams for (a) the function ${\cal K}$
and (b) the function ${\cal G}$ associated with the $B$ meson.
\vskip 0.5cm

\noindent
{\bf Fig. 7.} Dependence of (a) $f_1$ and (b) $f_2$ on the cutoff $b_c$
derived from $\phi_B^{(I)}$ (solid lines) and from
$\phi_B^{(II)}$ (dashed lines) for (1) $\eta=0.3$,
(2) $\eta=0.4$, and (3) $\eta=1.0$.

\vskip 0.5cm

\noindent
{\bf Fig. 8.} Dependence of $R(\eta)$
on $\eta$ derived from (1) $\phi_B^{(I)}$ and from (2)
$\phi_B^{(II)}$ (solid lines). Results from the soft pion theorems are
also shown (dashed line).

\end{document}